\documentclass[aps,prb,twocolumn,superscriptaddress,showpacs,floatfix,amsmath,amssymb]{revtex4}
\usepackage{graphicx}
\usepackage{color}
\usepackage{xcolor,soul}
\usepackage[normalem]{ulem}

\DeclareMathAlphabet\mathbfcal{OMS}{cmsy}{b}{n}

\newcommand{\br}[0]{ {\bf r} }

\newcommand{\KS}[0]{ {\rm {\tiny \tiny KS}} }

\newcommand{\bJ}[0]{ {\bf J} }
\newcommand{\bA}[0]{ {\bf A} }
\newcommand{\bj}[0]{ {\bf j} }

\usepackage{color}
\usepackage{booktabs}
\usepackage{longtable}
\usepackage{bm}

\newcommand{\mr}[1]     {\ensuremath{\mathrm{#1}}}

\newcommand{\Hxc}[0]{  {\rm Hxc}  }

\newcommand{\x}[0]{  {\rm x}  }
\newcommand{\xc}[0]{ {\rm xc}  }

\newcommand{\Ex}[0]{ E_\x }

\newcommand{\Exc}[0]{ E_\xc }

\newcommand{\Ec}[0]{ E_{ {\rm c} }}

\definecolor{Mygrey}{gray}{0.80}
\definecolor{lteal}{rgb}{0.10,0.60,0.70}
\definecolor{dkred}{rgb}{0.80,0.10,0.00}

\newcommand{\comment}[1]{}

\begin{document}

\title{Meta-Generalized-Gradient Approximation made Magnetic}
\author{Jacques K. Desmarais}
\email{jacqueskontak.desmarais@unito.it}
\affiliation{Dipartimento di Chimica, Universit\`{a} di Torino, via Giuria 5, 10125 Torino, Italy}

\author{Alessandro Erba}
\affiliation{Dipartimento di Chimica, Universit\`{a} di Torino, via Giuria 5, 10125 Torino, Italy}

\author{Giovanni Vignale}
\affiliation{Institute for Functional Intelligent Materials, National University of Singapore, 4 Science Drive 2, Singapore 117544}

\author{Stefano Pittalis}
\email{stefano.pittalis@nano.cnr.it}
\affiliation{Istituto Nanoscienze, Consiglio Nazionale delle Ricerche, Via Campi 213A, I-41125 Modena, Italy}

\date{\today}

\begin{abstract}
The Jacob's ladder of density functional theory (DFT) proposes the
compelling view  that by extending the form of successful approximations --- being guided by exact conditions and  selected  (least empirical) norms ---
upper rungs will do better than the lower, thus allowing to balance accuracy and computational effort.
Meta-generalized-gradient-approximations (MGGAs) belong to the last rung of the semi-local approximations before hybridization with non-local wave function theories. 
Among the MGGAs, the Strongly Constrained and Appropriately Normed Approximation (SCAN) greatly improves upon GGAs from the lower rung. But the over magnetized solutions of SCAN  make GGAs more reliable for magnetism. 
Here, we provide a solution that 
satisfies the most pressing {\em desiderata} for a density functional approximations
for ferromagnetic, antiferromagnetic and non-collinear states. The approach is  available in an implementation in the \textsc{Crystal} electronic structure package.
\end{abstract}


\maketitle

{\em Introduction.}~
Nearly one hundred years after Heisenberg proposed the first quantum mechanical explanation of ferromagnetism\cite{heisenberg1928theorie} magnetism continues to pose a challenge for \textit{ab initio} electronic structure theory. An unsettling example is given by the recently developed strongly-constrained and appropriately normed (SCAN) meta-generalized gradient
approximation (MGGA) for density functional theory (DFT).\cite{SCAN,CHEMSCAN}  

SCAN satisfies a large set of exact constraints, from known properties of the exact exchange-correlation (xc) functional and gives accurate results 
at a computational cost of about a generalized-gradient-approximation (GGA). Thus,
it can be considered a modern milestone from which it is possible to develop more accurate approximations.\cite{TASK,LAK}

But SCAN does {\em not} reproduce correctly the stability and properties of prototypical ferromagnets (FM) like Fe.
\cite{Fu2018,Ekholm2018,Fu2019,Trickey2019,Tran2020} 
Overestimated tendency toward magnetism and exaggeration of magnetic energies are obtained for transition metal systems.
Extension of the analysis  to antiferromagnetic (AFM) solids has shown that modifications of MGGAs yielding an improved description of FM metals produce a deterioration in AFM insulators, and vice versa.\cite{Tran2020} The status is such that MGGAs yield {\em no} advantage over  lower rungs of the semi-local approximation.\cite{Romero2018,Tran2020}

Here, we argue that the problem has persisted because the most
important exact constraint
for dealing with magnetism 
has not guided  most of the DFT developments thus far. We resolve this issue  by switching from collinear spin-DFT\cite{BarthHedin:72,GL76} (the current standard methodology) to the fully non-collinear (nc) Vignale-Rasolt-Bencheikh spin-current DFT  (SCDFT) -- a complete DFT theory for dealing with magnetism within the framework of the two-component Pauli equation.\cite{VignaleRasolt:88,Bencheikh:03} As will be shown, this extended framework can also solve problems for {\it collinear} DFT.

In SCDFT,  the most distinctive property
of the exact xc energy functional  is the
form invariance under U(1)$\times$SU(2) gauge transformations.\cite{VignaleRasolt:88,Bencheikh:03,Pittalis2017} 
These transformations rotate spinors as well as their phase  at each point in space  [see Eq. \eqref{TGAUGE} below]. 
Failing to satisfy this invariance leads to  functional forms that do not necessarily distinguish between physical variations of the spin density and those that arise from a mere change of reference frame. 
The gauge invariance of the exact xc-energy functional is thus the most important exact constraint for guiding the construction of magnetic approximations.

Working  within the general SCDFT,  extra  building blocks  for the construction of  extended approximate functional forms can be gained.\cite{Pittalis2017} In this work, we shall see, the collinear isoorbital indicators used in the SCAN (more below), which  treat   spin-up and  spin-down channels {\em separately}, yield  unbalanced magnetic solutions, \cite{Trickey2019,Tran2020} and can be replaced by a unified non-collinear quantity normed by the gauge principle. As a result, the major issues observed in the collinear SCAN solutions get resolved, and the revised functional form is seamlessly extended to the more general  non-collinear regime.

{\em From DFT to collinear Spin-DFT.~}
DFT can in principle  find the   particle density and total energy of  real {\em interacting}  ground states by solving 
(self-consistently) an auxiliary {\em non}-interacting problem 
\begin{eqnarray}
\label{KSE}
\left\{ -\frac{1}{2} \nabla^2 +  v_{\rm ext}  + \int d^3r' \frac{n(\br')}{|\br -\br'|} + v_{\xc} \right\} \varphi_{i}  = \varepsilon_{i} \varphi_{i}  \; ,
\end{eqnarray}
which is the celebrated Kohn-Sham equation. This is a Schr\"odinger-like equation that describes single-particle orbitals corresponding to  the actual external potential (the second term on the l.h.s.) added to  the
Hartree potential (the third term on the l.h.s.) and the xc potential, $v_{\rm xc}$. 
The latter potential is such that the actual particle density is retrieved
$n = \sum^{occ}_i |\varphi_{i}|^2 $.~\cite{noteION}
Correspondingly, the energy of the ground state of the real system $E$ is thus obtained by 
 \begin{eqnarray}\label{Etot}
E =   T_s +  \int d^3r~n ~ v_{\rm ext} +  E_{\rm H}[n] + \Exc[n] \; ,
\end{eqnarray}
where 
 $
T_s = \frac{1}{2} \sum_{i}^{\rm occ} \int d^3r~ |\nabla \varphi_{i}|^2
$
is the Kohn-Sham kinetic energy,
$E_{\rm H}[n] = \int \int d^3r d^3r' \frac{n(\br)n(\br')}{2|\br - \br'|}$ is the Hartree energy  and 
$\Exc[n]$ is the xc energy functional.  
For retrieving the exact total energy of the actual state, $\Exc[n]$  accounts for 
the electron-electron energy beyond the classical evaluation obtained via $E_{\rm H}[n]$  as well as the  kinetic energies beyond $T_s$.
Also notice that $v_{\rm xc}= \delta E_{\rm xc}/ \delta n$.
All that is needed to make the above approach practical is a density functional approximation (DFA) for $\Exc[n]$. 
Solutions to $n$ and $E$ can thus be found  in an iterative fashion.

The first and simplest   $E^{\rm DFA}_{\rm xc}[n]$ is the local density approximation (LDA)~\cite{KohnSham:65}
\begin{eqnarray}\label{LSDA}
 \Exc^{\rm LDA}[n] = \int d^3r~ n~ \epsilon^{\rm unif}_{\rm xc}(n) \; ,
\end{eqnarray}
where the xc energy density is approximated locally with that of the uniform gas, $\epsilon^{\rm unif}_{\rm xc}(n)$, for which analytical and accurate numerical reference data are available.
To capture the variations of more realistic states, a GGA uses
information retrieved from the  gradients of the density via $s=|\nabla n|/\left[2(3\pi^2 )^{1/3}n^{4/3}\right]$
\begin{eqnarray}\label{xPBE}
\Exc^{\rm GGA}[n] = \int d^3r~ n~ \epsilon^{\rm unif}_{\rm xc}(n) ~ F_{\rm xc}^{\rm GGA} (n,s) \;,
\end{eqnarray}
where $F_{\rm xc}^{\rm GGA} (n,s)$ is the xc enhancement factor that connects the LDA energy densities to the GGA energy densities. For slowly varying density, the LDA is retrieved (i.e. $F_{\rm xc}^{\rm GGA} (n,0)=1$).
The most popular choice for the enhancement factor is the one by Perdew, Burke and Ernzerhof (PBE) $F_{\rm xc}^{\rm PBE}$.   \cite{perdew1996generalized}

MGGAs~\cite{PKZB99,TPSS03,SCAN} introduce an extra dependence beyond the
electron density by employing 
\begin{eqnarray}\label{MGGA}
\Exc^{\rm MGGA}[n] = \int d^3r~ n~ \epsilon^{\rm unif}_{\rm xc}(n) ~ F^{\rm MGGA}_{\rm xc} (n,s,\tau) \;,
\end{eqnarray} 
where $\tau = 1/2\sum_i^{\mr{occ}}|\nabla \varphi_i|^2$ is the (KS) kinetic energy density. 
\cite{LL2022}
Here we focus on those MGGAs that depend on
$\tau$ via
\begin{eqnarray}\label{ALPHA}
\alpha := \frac{\tau - \frac{|\nabla n|^2}{8n}}{\tau_{\rm unif}} \; ,
\end{eqnarray}
where $\tau_{\rm unif} = \frac{3}{10} (3\pi^2)^{2/3}\, n^{5/3}$ 
is the kinetic energy density of the uniform gas at density $n$. In particular, the SCAN form is defined by setting $F_{\mr{xc}}^{\mr{SCAN}}(n,s,\alpha) = h_\mr{xc}^1 (n,s) 
+ \left[h_\mr{xc}^0 (n,s) - h_\mr{xc}^1 (n,s) \right]f_\mr{xc}(\alpha) 
$
where $f_\mr{xc}(\alpha)$ interpolates between the semi-local xc enhancement factors for single-orbital densities ($h_\mr{xc}^0$) and  for slowly-varying densities ($h_\mr{xc}^1$). This modeling exploits the fact that $\alpha$ [see Eq. \eqref{ALPHA} ] tends to zero in the iso-orbital limit (i.e., when the density is dominated by a single occupied orbital), tends to one in the uniform-density limit, and becomes  $\gg 1$ in regions of low density between closed shells

For solving ground states involving  collinear spin polarization either
induced by external Zeeman couplings or spontaneously via magnetic phase transitions, the standard approach is to switch from DFT to  Spin-DFT.\cite{BarthHedin:72,GL76}
The method requires to solve a spin-dependent, single-orbital problem
\begin{eqnarray}
\label{GKS-RES}
\left\{ -\frac{1}{2} \nabla^2 +  v_{{\rm ext},\sigma}  + v_{\Hxc,\sigma} \right\} \varphi_{i,\sigma}  = \varepsilon_{i,\sigma} \varphi_{i,\sigma} \;.
\end{eqnarray}
Here, $E_{\rm H}[n]$ and corresponding $v_{\rm H}$ are unchanged but the spin-dependent xc-potential, $ v_{\xc,\sigma} = \delta \Exc/\delta n_\sigma$, is generated  from the  functional  $\Exc[n_\uparrow,n_\downarrow]$, which depends both on $n_\uparrow$ and $n_\downarrow$ and not just on  their sum $n=\sum_\sigma n_\sigma$.

Extended forms for exchange can be generated  from given DFT approximations using the  {\it collinear} spin decomposition for exchange~\cite{noteSCALE}
\begin{eqnarray}\label{SpSc}
 \Ex^{\rm SDFA}[n_\uparrow,n_\downarrow] = \frac{1}{2} \Ex^{\rm DFA}[2n_\uparrow] + \frac{1}{2} \Ex^{\rm DFA}[2n_\downarrow]\;.
\end{eqnarray}
For correlation, we must use
\begin{eqnarray}\label{cDFA}
\Ec^{\rm SDFA}[n_\uparrow,n_\downarrow] = \int d^3r~ n~ \varepsilon^{\rm unif}_{\rm c}(n,\zeta) ~ F^{\rm SDFA}_{\rm c} (n,\zeta,s,\alpha) \;.
\end{eqnarray}
Unlike exchange, the correlation functional depends in a non separable fashion on $n_\uparrow$ and $n_\downarrow$ through the spin polarization (density) $\zeta=\vert (n_\uparrow-n_\downarrow)\vert /(n_\uparrow+n_\downarrow)$;
$\alpha = (\tau - \frac{|\nabla n|^2}{8n})/\tau_{\rm unif}d_s$ is the spin-summed iso-orbital indicator with 
$\tau = 1/2\sum_{i,\sigma}^{\mr{occ}}|\nabla \varphi_i|^2$ and $d_s=\frac{1}{2}[(1+\zeta)^{5/3}+(1-\zeta)^{5/3}]$. Also note that the exchange functional of Eq. \eqref{SpSc}, $\Ex^{\rm DFA}[n_\uparrow,n_\downarrow]$, can be equivalently reexpressed as $\Ex^{\rm DFA}[n,\zeta]$ in terms of the total density $n$ and polarization $\zeta$.

{\em Issues.}~ Notice that $\alpha$ does {\em not} carry any explicit information on $\nabla \zeta$. Therefore, the above expressions for GGAs {\em and} MGGAs come with a striking asymmetry. While the gradients of $\zeta$  are included in the exchange functional indirectly via the spin-scaling relation [i.e., Eq. \eqref{SpSc}], the correlation form has {\em no} dependence on $\nabla \zeta$.

This asymmetry reflects the lack of invariance of the approximate forms  under U(1)$\times$SU(2) gauge transformations.  Although this deficiency might not be perceived as a problem as long as we consider scenarios that are not gauge symmetric  --- such as  FMs or AFMs --- one should not confuse the symmetry properties of {\em particular} solutions with
those of the exact functional. To sort the latter out, we must enlarge our modeling framework.

{\em Switching to SCDFT.}~ Let us
account not only for a scalar multiplicative potential $v_{\rm ext}$  and a multiplicative magnetic field
${B}^a$, but also for a 
(particle-) vector potential $\bA$ and a spin-vector potential $\bA^a$.~\cite{noteNOTATION}
While $\bA$ is useful to represent the vector potential of an external magnetic field,
 $\bA^a$ is useful to represent the (one-body) spin-orbit couplings in the system.~\cite{VignaleRasolt:88,Bencheikh:03}

The KS equations in SCDFT
have the form of single-particle Pauli equations~\cite{VignaleRasolt:88,Bencheikh:03}
\begin{equation}\label{KSeq}
\left[ \frac{1}{2}\left( -i   \nabla + \frac{1}{c} { \mathbfcal{A} }_\KS  \right)^2 +  {\cal V}_\KS  \right] \Phi_k = \varepsilon_k \Phi_k \; ,
\end{equation}
where  $\Phi_k$ are two-component spinors,
$
{ \mathbfcal{A} }_\KS =\left( {\bf A} + {\bf A}_{{\rm xc}} \right) +  \sigma^a \left( {\bf A}^a + {\bf A}^a_{{\rm xc}} \right)
$,
$
{\cal V}_\KS = \left( v + v_{\rm H}+ v_{\rm xc} \right) + \sigma^a\left( B^a + B^a_{\rm xc} \right) 
+ \frac{1}{2c^2} \left[ \left(  {\bf A} + \sigma^a {\bf A}^a \right)^2-  {\mathbfcal A}^2_\KS \right]
$,
in which $\sigma^a$ denotes the component of the vector of Pauli matrices,
$
{\frac{1}{c}}{\bf A}_{\rm xc}= \delta E_{\rm xc}/\delta {\bf j}$
is an Abelian xc-vector potential,
$
{\frac{1}{c}}{\bf A}^a_{\rm xc} = \delta E_{\rm xc}/\delta { \bf J}^a
$
is the $a$-th component of a non-Abelian xc-vector potential,
$
{B}^a_{\rm xc} = \delta E_{\rm xc}/\delta {m^a}
$
is the $a$-th component of a xc-magnetic-like potential,
$
 v_{\rm xc} = \delta E_{\rm xc}/\delta n
$
is a xc-scalar potential, and $v_{\rm H}$ is the usual Hartree potential.
The KS densities are
obtained from the (occupied) two-component KS spinors as follows:
$n = \sum_{k}^{\rm occ} \Phi^\dagger_k \Phi_k$,
${\vec m} = \sum_{k}^{\rm occ} \Phi^\dagger_k\;{\vec \sigma}\; \Phi_k
$, $
{\bf j} = \frac{1}{2i} \sum_{k}^{\rm occ} \Phi^\dagger_k \left[ \nabla \Phi_k\right] -  \left[ \nabla \Phi^\dagger_k \right] \Phi_k$,
and ${\vec {\bf J}} = \frac{1}{2i} \sum_{k}^{\rm occ} \Phi^\dagger_k {\vec \sigma}  \left[ \nabla \Phi_k \right] -  \left[ \nabla \Phi^\dagger_k \right] {\vec \sigma} \Phi_k
$.

Because  $E_{\rm xc}[n, {\vec m}, {\bf j}, {\vec {\bf J}}]$ must yield the {\em exact} interacting densities via the KS densities, it must also be form invariant under (local) U(1)$\times$SU(2) gauge transformations.\cite{VignaleRasolt:88,Bencheikh:03,Pittalis2017} In detail, such a transformation acts on spinors as follows
\begin{equation}\label{TGAUGE}
{\Phi}' ({\bf r}) = \exp \left[ \frac{i}{c}  \chi({\bf r}) {\mathrm I} + \frac{i}{c} {\lambda^a}({\bf r})  {\sigma^a} \right]  {\Phi} ({\bf r})\;,
\end{equation}
where I is the $2\times2$ identity and $\chi$ and $\lambda^a$ are arbitrary real-valued functions.

We recall from Eq. \eqref{MGGA}  that MGGAs, such as SCAN, explicitly depend on the KS kinetic energy density $\tau$. For dealing with nc solutions, the kinetic energy density  $\tau$  must be expressed in terms of the occupied two-component KS spinors  
$\tau_{\rm nc} = 1/2\sum_k^{\mr{occ}}  \Big(  \partial_\mu \Phi^{\dagger}_k \Big) \Big( \partial_\mu \Phi_k \Big)$.
But, it can be directly verified that $\tau_{\rm nc}$ is not gauge invariant.\cite{Pittalis2017} Thus, $\alpha_{\rm nc}$ obtained 
by replacing $\tau$ with $\tau_{\rm nc}$ in Eq. \eqref{ALPHA} is not invariant either.
The corresponding extension of the SCAN functional by letting $\alpha \to \alpha_{\rm nc}$ would therefore lack the most fundamental property for dealing with magnetism. Fortunately, this  problem can be  fixed  by the replacement $ {\tau}_{\rm nc} \rightarrow \widetilde{\tau}_{\rm nc}$ with\cite{Pittalis2017,zwoelf}
\begin{equation}\label{tTAU}
 \widetilde{\tau}_{\rm nc}
=   
 \tau_{\rm nc}  +   \frac{{m}^a {\tau}_{\rm nc}^a}{n} 
 +  \frac{\nabla {m}^a \cdot \nabla {m}^a  }{8 n}  - \frac{\bj \cdot \bj}{ 2n} - \frac{ \bJ^a \cdot \bJ^a  }{ 2n  } \;,
\end{equation} 
where 
$
{\tau}^{a}_{\rm nc} = 1/2 \sum_{k}^{\rm occ} \Big(  \partial_\mu \Phi^{\dagger}_k \Big) {\sigma}^{a} \Big( \partial_\mu \Phi_k \Big)\;
$
is the spin-kinetic energy density.

Eq.~\eqref{tTAU} not only includes $\nabla \vec{m}$ and the (spin-)paramagnetic currents but, for consistency with the exact gauge-invariance requirements, also   the coupling between ${\vec m}$ and ${\vec \tau}_{\rm nc}$.

Eq.~\eqref{tTAU} includes the well-known particle current corrections that ensure the U(1) invariance of the MGGA, which has been advanced both in practice and in theory.\cite{Dobson93,Becke-j02,TP05,Pittalis2006,Pittalis07,Rasanen09,furness2015current,tellgren2018uniform,laestadius2019kohn,pemberton2022revealing}
It also includes the  spin current corrections that were recently shown to ensure the SU(2) invariance of the MGGA  for  time-reversal symmetric spin-orbit coupling --- the JSCAN.\cite{JSCAN}
Eq.~\eqref{tTAU} was also at the center of the derivation of a non-collinear ELF.\cite{zwoelf} But 
it did not find exploitation in energy functional approximations to determine magnetic solutions.

We recall that the problems of overmagnetization by SCAN 
 were imputed, in Refs.~\onlinecite{Trickey2019,Tran2020}, to the different behaviour of the isoorbital indicators in the exchange of SCAN for the two spin channels. Below, we solve these problems by invoking a single $\widetilde{\alpha}$ for both spin channels --- normed by the  gauge principle. For capturing energies rather than just bond pictures, the following gauge invariant extension of SCAN suggests itself

\begin{subequations}
\begin{equation}\label{mJSCANx}
E^{\rm nc}_{\rm x} [ n,\vec{m},\mathbf{j},\vec{\mathbf{J}} ] =  \int d^3r ~ n ~ \varepsilon^{\rm unif}_{\rm x} \left( n,\zeta_{\rm nc} \right) ~ F^{\rm SCAN}_{\rm x} \left( n, s , \widetilde{\alpha}_{\rm nc} \right)\;,
\end{equation}
and
\begin{equation}\label{mJSCANc}
E^{\rm nc}_{\rm c} [ n,\vec{m},\mathbf{j},\vec{\mathbf{J}} ] = \int d^3r~ n~ \varepsilon^{\rm unif}_{\rm c}(n, \zeta_{\rm nc}) ~ F^{\rm SCAN}_{\rm c} (n,\zeta_{\rm nc},s,\widetilde{\alpha}_{\rm nc}) \;,
\end{equation}
\end{subequations}
where $\zeta_{\rm nc} \equiv \vert \vec{m} \vert/n$, and
\begin{equation}\label{eqn:alphanc}
\widetilde{\alpha}_{\rm nc} \equiv \frac{ 2n \widetilde{\tau}_{\rm nc}- \frac{ ( \nabla n  ) \cdot  ( \nabla n  )}{4 } }{  \left( n+\vert \vec{m} \vert \right) \tau_{\rm unif}^{+} +   \left( n-\vert \vec{m} \vert \right) \tau_{\rm unif}^{-}  }
\end{equation}
in which $\tau_{\rm unif}^{\pm } \equiv 3/10(3\pi^2)^{2/3}(n\pm \vert \vec{m} \vert)^{5/3}$, and all {\em uniform} gas quantities have been upgraded to an nc form by locally quantizing spin along the axis defined by the spin density $\vec{m}$.\cite{KueblerWilliams:88,noteISO}
The derivation of Eqs. \eqref{eqn:alphanc}  and \eqref{tALPHA} is provided in the Supplementary Material (this includes Ref.s \onlinecite{GiulianiVignale,oliver1979spin,Lehtola2020,desmarais2021spin,vilela2019bsse,peterson2007energy,dolg2005improved,balabanov2005systematically,wilson1996gaussian,campetella2020hybrid,saunders1992electrostatic,towler1996density,doll2001analytical,doll2006analytical,desmarais2023efficient,dachsel1999multireference,yang2018structure,wenzel2022effect,el1991magnetic,jones2022origin,huber1979constants}).\cite{ESI,noteCOMP}

For self-consistent {\em collinear} solutions, a {\em modified} SCAN functional (named mSCAN) is straightforwardly obtained solely in terms of $n^\sigma$,  $\varepsilon^{\rm unif}_{\rm x/c}(n^{\uparrow}, n^{\downarrow})$, and
\begin{eqnarray}\label{tALPHA}
\widetilde{\alpha} \equiv 
\frac{n^\uparrow \tau^\uparrow+ n^\downarrow \tau^\downarrow - \frac{\left( \nabla n^\uparrow \right) \cdot \left( \nabla n^\downarrow \right)}{4}}{ n^\uparrow \tau^\uparrow_{\rm unif}+ n^\downarrow \tau^\downarrow_{\rm unif} }\;,
\end{eqnarray}
where
\and $\tau^\sigma_{\rm unif}=3/10(6\pi^2)^{2/3}(n^\sigma)^{5/3}$.\cite{noteHolzer2022}
Finally, the mSCAN reduces to the SCAN on closed-shell solutions (more below).

{\em  Notable magnetic SCAN failures --- resolved.}
Let's scrutinize the novel collinear solutions.
We implemented the mSCAN functional in a developer's version of the \textsc{Crystal23} program.\cite{erba23cry}  We study current-less, collinear magnets. Thus, the single-particle equation of SCDFT given in Eq. \eqref{KSeq} is considerably simplified, as the vector potentials ${\bf A}_{\rm xc}$ and ${\bf A}^a_{\rm xc}$ vanish,  and we must self-consistently solve the simpler Eq. \eqref{GKS-RES} in a basis of Bloch orbitals $\varphi_{i,\sigma,\mathbf{k}}$ (with wavevector $\mathbf{k}$) formed from a linear combination of atom-centered spherical Gaussian-type functions (GTF).\cite{desmarais2018generalization} 
In Eq. \eqref{GKS-RES} for mSCAN, in a generalized KS context,\cite{Seidl1996} the spin-dependent xc potential $v_{{\rm xc},\sigma}$ includes an additional semilocal term coming from the spin-kinetic energy dependence.\cite{Neumann1996,Lehtola2020,desmarais2024generalized} Computational details for the general case are provided in the SM.

Before applying mSCAN to prototypical magnets, notice that the x-mSCAN does not fulfill the (strictly collinear) 
spin decomposition relation of Eq. \eqref{SpSc}, yet it fulfills the exact linear scaling conditions (collinear or non-collinear). An extended analysis is provided in the SM.

Although SCAN fits the energy of the H atom as an appropriate norm and  assigns no correlation to  one-electron systems,  Hartree and x-SCAN do not cancel for one-electron systems. mSCAN  breaks both the constraint of zero correlation energy and zero Hx for one-electron systems. Yet, tests show that the overall treatment of the Hxc energies in mSCAN  is nonetheless appropriately normed by fulfilling the gauge invariance --- not fulfilled  by the SCAN form. For example, for H$_2^+$ (see Figure S1) neither SCAN nor mSCAN matches the exact energies:  while SCAN is accurate at short separation, mSCAN is  accurate at large intermolecular separation.  Taking another one-electron example, the potential energy curve of Li$_2^{5+}$ is most accurate (as compared to LDA, GGA and SCAN) by mSCAN at {\em all} distances (Fig. S2).

\begin{table}[hb!]
\caption{Spin moments (in $\mu_b$/atom) and, within parentheses, magnetic energies (in meV) of FM metals as calculated with DFAs and compared against experiments.}
\label{tab:ferro}
\vspace{2pt}
\begin{tabular}{rccccc}
\hline
\hline
&     Fe & Co & Ni \\
&&\\
LDA & 2.23 (467)& 1.57 (396)& 0.56 (40.8)\\
PBE & 2.26 (581)& 1.59 (504)& 0.56 (49.0)\\
SCAN& 2.62 (1177)& 1.74 (1175)& 0.66 (113)\\
mSCAN & 2.15 (461) & 1.58 (447)& 0.56 (47.3)\\
&&\\
EXP & 1.98-2.08\text{\cite{chen1995experimental,reck1969orbital,scherz2004spin}} & 1.52-1.62\text{\cite{chen1995experimental,reck1969orbital,moon1964distribution,scherz2004spin}} & 0.52-0.55\text{\cite{reck1969orbital,scherz2004spin}}\\
\hline
\hline
\end{tabular}
\end{table}

Next, we consider also materials. In  Table \ref{tab:ferro} we list the spin magnetic moments (calculated analytically in the Gaussian basis set) and the magnetic energies (in meV), calculated as energy differences w.r.t. the self-consistent spin-restricted solution for the FM metals Fe, Co and Ni. All calculations employ the experimental geometries reported in Ref. \onlinecite{Tran2020}, unless explicitly stated otherwise. Comparisons are provided against the LDA (employing the parametrization of correlation energies by Vosko, Wilk and Nusair, and the usual Slater-Dirac energy functional for exchange SVWN5),\cite{vosko1980accurate} the GGA of PBE,\cite{perdew1996generalized} and the original spin-unrestricted SCAN.\cite{SCAN} Comparison against experimental data for the spin-magnetic moments are also given.

In accord with previous reports,\cite{Fu2018,Tran2020,Trickey2019}
we find an agreement with experiments at the level of LDA, which is not significantly changed with the GGA. SCAN displays the previously reported  overmagnetization in all FM metals, both in its significant overestimation of the magnetic moments, as well as overestimation of the magnetic energies.\cite{Fu2018,Tran2020,Trickey2019} In the case of Fe, for example, the self-consistently predicted magnetic moment of 2.62 $\mu_b$ by SCAN represents an overestimation of around 30 \%. Overmagnetization of around 10 \% and 15 \% w.r.t. experimental values are predicted by SCAN for Co and Ni respectively, while the LDA and GGA both fall within the experimental uncertainty. Magnetic energies of SCAN are consistently more than double the values predicted by the LDA and GGA for the FM metals. The problem is completely solved by switching to our gauge-invariant mSCAN approach, which provides similar magnetic moments (and magnetic energies) as the LDA and GGA on Co and Ni, while the predicted magnetic moment on Fe of 2.15 $\mu_b$ is the closest to the experimental range of 1.98-2.08 $\mu_b$.

The improvement provided by our approach is even more striking when applied to the non-magnetic metals V and Pd (see Table \ref{tab:nonferro}), as well as the difficult case of Cr$_2$.\cite{larsson2022chromium,zhang2020localization} We perform the calculations starting from a spin-polarized guess and let the spin moments relax during the self-consistent procedure. Like in previous works,\cite{Fu2018,Tran2020} SCAN calculations incorrectly predict a net magnetic moment of 0.33 and 0.44 $\mu_b$ on V and Pd respectively. In the case of Pd, the incorrect magnetic behaviour is also predicted by  PBE. For Cr$_2$, the LDA, PBE and SCAN all predict a spurious magnetization at the equilibrium geometry (ground state properties are reported in Table S1 of the SM). Remarkably, the correct non-magnetic behaviour of V, Pd and Cr$_2$ is predicted by the proposed gauge-invariant mSCAN approach. As for itinerant two-dimensional systems, in Graphene, SCAN incorrectly predicts a magnetic insulating phase,\cite{zhang2020localization} while mSCAN correctly provides a non-magnetic, metallic  state, with a Dirac point at the zone corners (Table S2).

\begin{table}[ht!]
\caption{Same as in Table \ref{tab:ferro} but for non-magnetic systems.}
\label{tab:nonferro}
\vspace{2pt}
\begin{tabular}{cccc}
\hline
\hline
&     V &Pd & Cr$_2$\\
&&\\
LDA & 0.00 (0.00)& 0.00 (0.00)& 2.69 (0.30)$^\ast$\\
PBE & 0.00 (0.00)& 0.20 (0.70)& 5.32 (0.87)$^\ast$\\
SCAN& 0.57 (6.27)& 0.44 (17.2)& 5.62 (10.3)$^\ast$\\
mSCAN & 0.00 (0.00)&0.00 (0.00)& 0.00 (0.00)\\
&&\\
EXP & 0.00\text{\cite{weber1997magnetism}} &0.00\text{\cite{kudasov2007surface}}& 0.00\text{\cite{larsson2022chromium}} \\
\hline
\hline
\end{tabular}
\footnotetext{$^\ast$ Values in eV.}
\end{table}

Although the failures of SCAN on itinerant magnets are daunting, accurate magnetic moments and gaps have been predicted in the AFM Mott insulators FeO, CoO and NiO, which exhibit more localized magnetism.\cite{zhang2020symmetry,Tran2020} Values obtained from our mSCAN on these highly-correlated Mott insulators are provided in Table \ref{tab:anti}, including a comparison against the experiment both on magnetic moments and fundamental band gaps (in parenthesis, given in eV). The LDA and GGA incorrectly predict a metallic behaviour for FeO and CoO, while both SCAN and mSCAN provide the qualitatively correct insulating behaviour. All functionals provide similar values for the spin magnetic moments in the Mott insulators, although both moments and band gaps are larger with SCAN. We also consider a topological AFM -- the Chern magnet TbMn$_6$Sn$_6$, in which calculation of spin moments is a reported challenge for SDFT.\cite{annaberdiyev2023role} With mSCAN, we obtain excellent agreement against the experimental values (with  large improvement over SCAN and PBE) in Table S3.

\begin{table}[hb!]
\caption{
Spin moments (in $\mu_b$/atom) and, within parenthesis, band gaps (in eV) of AFM Mott insulators.} 
\label{tab:anti}
\vspace{2pt}
\begin{tabular}{rccccc}
\hline
\hline
&  FeO & CoO & NiO \\
&&\\
LDA & 3.51 (0.000)& 2.40 (0.000)& 1.20 (0.452)\\
PBE & 3.57 (0.000)& 2.48 (0.000)& 1.37 (0.986)\\
SCAN& 3.67 (0.635)&  2.65 (1.29)& 1.62 (2.66)\\
mSCAN & 3.57 (0.224)& 2.52 (0.440)& 1.44 (1.29)\\
&&\\
EXP$^\ast$ & 2.32-4.00\text{\cite{roth1958magnetic,battle1979magnetic,fjellvaag1996crystallographic}}  & 1.75-2.98\text{\cite{roth1958magnetic,khan1970magnetic,herrmann1978equivalent,jauch2001crystallographic}}  & 1.45-1.90\text{\cite{roth1958magnetic,fernandez1998observation,cheetham1983magnetic,neubeck1999observation}} \\
EXP & (2.40)\text{\cite{bowen1975electrical}} & (3.10-4.10)\text{\cite{gvishi1970transition}} & (3.7)\text{\cite{ksendzov1965forbidden}} \\
\hline
\hline
\end{tabular}
\footnotetext{$^\ast$We remove the orbital contributions provided in Ref. \onlinecite{Tran2020} from the experimental total moments (see caption of Table VI of Ref. \onlinecite{Tran2020}).} 
\end{table}

Next, we investigate the performance of mSCAN on structural optimizations in all tested systems (see Table S4). A previous attempt at partially fixing the spin moments of SCAN by varying the $c_{1x}$ parameter has produced significantly worsened results for lattice constants, with errors increased by around 50\%.\cite{Tran2020} Here,
mean absolute errors (MAE) decrease in the order LDA, GGA, mSCAN (SCAN), with values of 0.663, 0.521, 0.295 (0.319), which is again consistent with the principle of the Jacob's ladder of DFT.

Let us now scrutinize the dissociation of first-row diatomic molecules involving open-shell atoms.\cite{noteGUESS} Fig. S3 reports the case of  H$_2$ compared to the accurate result of Ref. \onlinecite{kolos1968}: the mSCAN and SCAN energies match up until an intermolecular separation of about 3 \AA, at which point a {\em spurious} magnetization appears in the  spin-unrestricted LDA, PBE and SCAN solutions. mSCAN provides a correct vanishing spin magnetization up to about 3.5 \AA.
In the C$_2$ molecule LDA, PBE and SCAN provide spurious magnetization even at the equilibrium geometry, which is resolved by mSCAN, see Fig. S4 and Table S5. mSCAN also provides an improvement over SCAN, with respect to the reference total energies of Ref. \onlinecite{booth2011breaking} on C$_2$, both at the equilibrium geometry and at large distances.
Fig. S5 reports the dissociation of the N$_2$ molecule.
mSCAN provides a significant improvement over LDA, PBE and SCAN at distances larger than about 1.5 \AA, where strong correlation becomes important (as also corroborated by comparison of multi- and single-reference coupled-cluster treatments).\cite{li2008full} 

A word of caution and of encouragement are in order: The novel functional behaves differently  from SCAN not only for  magnetized states but also for open-shell states, therefore atomization energies
should be scrutinized with care. We have also pointed out that band gaps may be worsened.
Therefore, superseding the SCAN functional in all respects will require further efforts, which, we believe, can exploit the recently discovered extra-degrees of freedom that are allowed by the MGGAs. \cite{LAK}

\begin{figure}[ht!!]
\centering
\includegraphics[width=6.3cm]{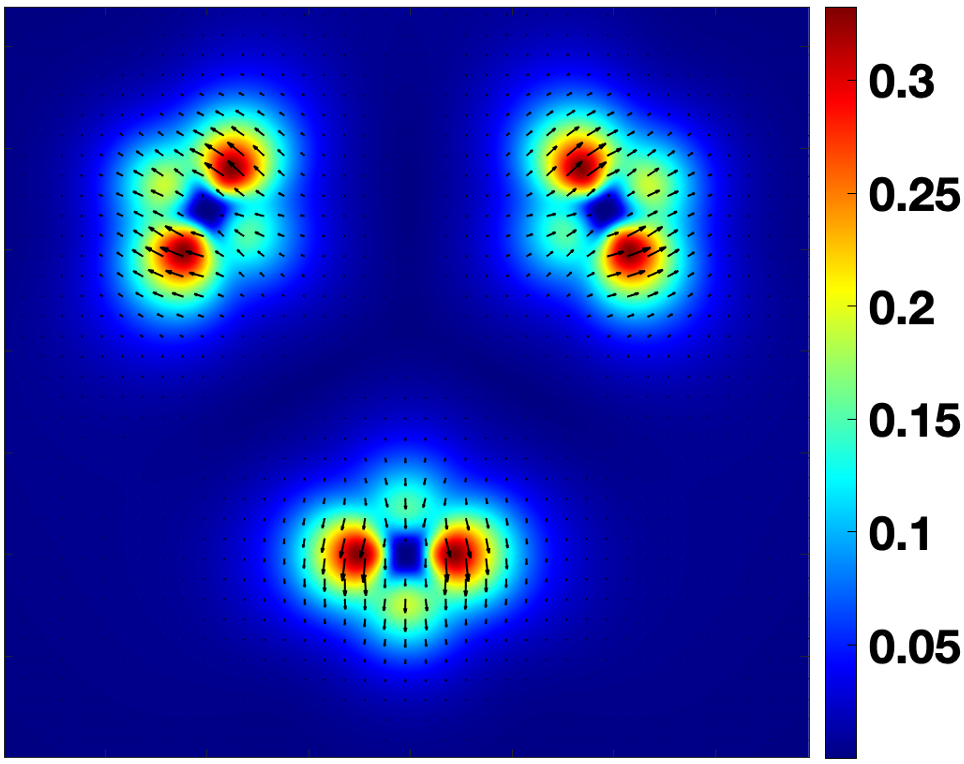}
\caption{Magnetization in $\mu_b$ of the Cr$_3$ molecule with the nc SCAN functional.}
\label{fig:Cr3}
\end{figure}

{\em Noncollinear extension.}~ 
Fully self-consistent SCDFT MGGA calculations for non-collinear magnetic states are reported in a parallel work.\cite{huebsch2025capturing} Ref. \onlinecite{huebsch2025capturing} also calls for a gauge-invariant nc-extension of SCAN.

Thus, we finally solve Eq. \eqref{KSeq} using the nc forms in Eq.s \eqref{mJSCANx} and \eqref{mJSCANc}. Technical details are provided in the SM, where we also demonstrate that the functional provides a non-zero xc magnetic torque, which vanishes over the full space. Last but not least, spin-orbit coupling can be accounted for self-consistently. So, our proposal fulfills the most pressing {\em desiderata} for a nc approach.

We applied it to the Cr$_3$ (bond length of 3.7 Bohr), which has been an important model system in previous studies.\cite{ScalmaniFrisch:12,zwoelf,huebsch2025capturing,tancogne2023,pu2023noncollinear} The magnetization reported in Fig. \ref{fig:Cr3} is indicative of a spin-frustrated system. We observe a spin-moment of 1.38 $\mu_b$, which is significantly lower than the SCDFT-MGGA value of 3.15  $\mu_b$ reported in Ref. \onlinecite{huebsch2025capturing}, but closer to the LDA value of 1.68 $\mu_b$.

{\em Conclusions. }~ 
While it remains a challenge to assess whether a semi-local density functional may be derived to embrace {\em all} the relevant 
non-magnetic and magnetic (both collinear and non-collinearly polarized) exact constraints, the core findings of this work solves  most urgent  failures for magnetism and resumes progress towards improved density functional approximations.

\begin{acknowledgments}
This research has received funding from the Project CH4.0 under the MUR program ``Dipartimenti di Eccellenza 2023-2027'' (CUP: D13C22003520001). GV was supported by the Ministry of Education, Singapore, under its Research Centre of Excellence award to the Institute for Functional Intelligent Materials (I-FIM, project No. EDUNC-33-18-279-V12).
\end{acknowledgments}

\appendix


\begin{thebibliography}{99}
\bibitem[{not({\natexlab{a}})}]{noteION}
\bibinfo{note}{For simplicity the expressions we assume integer occupation
  numbers but in applications fractional occupation may also occur.).}
\bibitem[{not({\natexlab{b}})}]{noteSCALE}
\bibinfo{note}{We refrain from using the term ``spin scaling relations'' to
  avoid confusion between aspects of the spin separability with aspects of
  uniform coordinate scaling in the more general non-collinear framework.}
\bibitem[{not({\natexlab{c}})}]{noteNOTATION}
\bibinfo{note}{We denote with bold characters, $\boldsymbol{v}$, quantities
  with spatial indices (Greek lower indices, $v_\mu$, when written explicitly);
  and use an arrow, $\vec{v}$, to denote quantities with spin indices (upper
  Latin indices, $v^a$, when written explicitly).}
\bibitem[{not({\natexlab{d}})}]{noteISO}
\bibinfo{note}{We stress that $\widetilde{\alpha}_{\rm nc}$ differs from a
  previously reported noncollinear iso-orbital indicator underlying the recent
  non-collinear extension of the ELF.\cite{zwoelf} This is due to the fact
  that, in the present context by reusing the LSDA as the cornerstone of the
  new approximation, we must use as a reference a uniform gas that is polarized
  non-collinearly.}
\bibitem[{not({\natexlab{e}})}]{noteCOMP}
\bibinfo{note}{For completeness, the collinear limit of Eqs. 13a and 13b is
  reported in the SM.}
\bibitem[{not({\natexlab{g}})}]{noteGUESS}
\bibinfo{note}{All calculations were performed with a spin-polarized starting
  guess from an antiferromagnetically coupled superposition of atomic Hartree
  Fock spin-densities.}
\end{thebibliography}

\end{document}